# NbN single-photon detectors with saturated dependence of quantum efficiency

Konstantin Smirnov[1,2,3], Alexander Divochiy[1,2], Yury Vakhtomin[1,2], Pavel Morozov[1], Philipp Zolotov[1,2], Andrey Antipov[1,2], Vitaliy Seleznev[1,2]

[1] Moscow State Pedagogical University, Moscow 119991, Russia
[2] JSC «Superconducting Nanotechnology» (SCONTEL), Moscow 119021, Russia
[3] National Research University Higher School of Economics, Moscow 101000, Russia

It was investigated the possibility of creating NbN superconducting single-photon detectors with saturated dependence of quantum efficiency versus normalized bias current. It was shown that the saturation increases for the detectors based on finer films with a lower value of $R_{s300}/R_{s20}$. The decreasing of $R_{s300}/R_{s20}$ related to increasing influence of quantum corrections to conductivity of superconductors and, in its turn, to decreasing electron diffusion coefficient. The best samples has constant value of system quantum efficiency 94% at $I_b/I_c \sim 0.8$ and wavelength 1310 nm.

**Introduction**

Over recent years, superconducting single-photon detectors (SSPD) have demonstrated continuous improvement of the key characteristics such as quantum efficiency [1-5], dark counts rate [6-8], jitter [9, 10], dead time [11-13] and noise-equivalent power [14]. Also SSPD has found numerous applications in various fields [15-18]. In order to increase absolute value of quantum efficiency (*QE*), to develop SSPD in the long wave infrared spectral region and to get quantum efficiency undependable on bias current, attempts were made to create SSPD with ultra-narrow superconducting strips of width 20-50 nm [19-21]. This approach revealed a number of drawbacks related mostly to increased technology requirements for producing such structures. Thus, it should significantly decrease the yield of fabricated SSPDs which demonstrate high quantum efficiency. Moreover, the critical current density of the superconducting structure usually drops at least proportionally with decreasing superconducting strip width leading to a significant reduction in the signal-to-noise ratio and increasing SSPD jitter value. Furthermore, the fabrication of detectors with narrow superconducting strips also requires increasing the superconducting strip length in order to keep fill-factor constant which leads to increasing its kinetic inductance and decreasing maximum count rate.

Another development direction of SSPD is the fabrication of single-photon detectors using material with a low-energy gap or a lower superconducting transition temperature $T_C$ which allows to reach a high quantum efficiency with saturated dependence of $QE(I_b)$, where $I_b$ is the detector bias current. For example, several materials such as WSi and MoSi [2, 14, 22, 23] were applied for SSPD fabrication. The authors [2, 23] explain that the SSPD can reach high (93 %) system quantum efficiency by using WSi due to its particular characteristics.

However, WSi-based detectors have a number of disadvantages. The time resolution is usually higher than 50 ps [23] which is probably due to low WSi critical temperature $T_C$ and low critical current $I_c$ [2]. Moreover, in order to reach a high system quantum efficiency for receiving system based on WSi detectors, the operating temperature should be below 1 K making this system complex and expensive.

In recent studies NbTiN was also proposed as an alternative polycrystalline material for SSPD characterized by a higher critical temperature and critical current density [24]. Based on recent results we believe that NbN and NbTiN possess similar properties related to the SSPD fabrication and chose NbN because of its deeply characterized properties. As it was shown before, NbN devices shows all desired characteristics when high optical absorption is provided. In case of waveguide SSPD on-chip detection efficiency of the device could be as high as 91%, timing resolution could be as low as 18 ps, alongside with high detection rates and ultra-low dark counts [25]. Although, coupling the source of radiation to such device usually leads to dramatic drop of system quantum efficiency and dark count rate growth.

The most promising and competitive fabrication technology of SSPD with high system quantum efficiency and tendency toward its saturation is the thickness reduction of an initial NbN superconducting film or fabrication of NbN thin film with a more disordered structure. In the article [26] authors made SSPDs only from 3.5 and 10 nm thick NbN films. Devices which were made from the thinner NbN films had higher intrinsic quantum efficiency. Also, [27] investigated SSPD detection efficiency for two different thicknesses NbN films. The main purpose of the current work is the investigation of NbN thin films over a wide range of thicknesses for fabrication of SSPD which have the dependence of $QE(I_b)$ with tendency toward its saturation in the range of bias currents close to critical current $I_c$. By measuring $R(T)$ for films with different thicknesses, we found that the residual-resistance ratios (RRR) changes significantly with the film thickness, and attempted to explain that this behaviour of the parameter RRR is due to the influence of quantum corrections to the conductivity. Next, we assume that the tendency toward to saturation of $QE(I_b)$ dependence with decreasing film thickness is not only due to decreasing cross sectional area of NbN strip, but also due to changed parameters of the film, which are confirmed by measurements of diffusion coefficients in films with the different thickness. In the conclusion we used multilayer antireflection coating for detectors which have pronounced saturated $QE(Ib)$ dependencies to achieve the highest possible system quantum efficiency (*SQE*). Similar studies of *R(T)* dependences for NbN films were previously performed in [25]. However the authors investigated the properties of epitaxial NbN films grown on MgO substrate and did not investigate its applicability for the manufacture of SSPD. Another significant difference between these films and our investigated films is that the authors [25] observed a positive temperature coefficient of resistance down to 3 nm film thickness. As well as increasing the plateau region we also attempted to reach the ultimate level of system quantum efficiency for our SSPD devices.

**Results and Discussions**

**SSPD fabrication.** The NbN films were deposited on a Si substrate with an additional Au/Si$_3$N$_4$ bilayer by reactive magnetron sputtering in the Ar and N$_2$ gas mixture. The main parameters of the obtained films are shown in Table 1. The measured and controlled parameters for solid films are superconducting transition temperature (in the range of 3.55–9.3 K, the thinnest film did not exhibit superconductive transition down to the temperature of 1.6 K), surface resistance (in the range of 450–4410 Ω/sq), film thickness (in the range of 2–9 nm). The thickness of the films was calculated by determined deposition rate of thick (~100 nm) NbN films measured using an atomic force microscope and investigated using x-ray photoelectron microscopy (XPS). The latter showed that deposited NbN films have several interlayers with different compositions and the increased total films thickness up to few nanometers compared to those are usually reported in the literature [28]. This result will be observed more closely in our future work. In view of obtained data we consider the values of $R_{s300}/R_{s20}$ (RRR) and Tc more significant than the thicknesses of the films by reason of more precise measurement techniques. The changes in the critical superconducting transition temperature and the film surface resistances were achieved by varying the film thickness. Ar and N$_2$ concentrations were kept constant for all films with different thicknesses, except of the small deviation for 8 nm-thick film, which led to a small increasing of $R_{300}$ with relatively to expected value. However, it did not influence general trend of RRR and $T_C$ shift.

We fabricated meander-shaped detectors [29] covering an area of 15x15 μm$^2$ based on films with the superconducting transition temperature of 8.1-9.3 K. The width of the superconducting strip is ~100 nm and the fill-factor is ~0.5. The deposition of an additional Au/Si$_3$N$_4$ layers on the Si substrate serve as an optical resonant structure which increases the absorption coefficient of SSPD active area up to 40-50%. Moreover, we deposited an antireflection coating (ARC) on top of several detectors that allowed us to increase the absorption coefficient up to ~98%. There were several articles, where authors consider ARC for SSPDs [2, 30]. We have been modeling a lot of different compositions of materials for ARC, such as MgO, MgF2, SiO2, Si, Ta2O5, TiO, Al2O3, ZnS and etc. As an ARC we used Al$_2$O$_3$/Si/Al$_2$O$_3$ three-layered structure. The best absorption coefficient for $\lambda$ = 1310 nm has been obtained for the structure Si-wafer/Au/Si3N4/NbN/Al$_2$O$_3$/Si/Al$_2$O$_3$. All layers mentioned above (Si$_3$N$_4$, Al2O3, Si and Au) were fabricated by the electron-beam evaporation method. The ARC was applied through a mask made of metal foil with holes of 0.5 mm diameter which was centered over the sensitive element of the detector.

**Measurements methods.** In order to measure the temperature dependences of resistance of superconducting film and detectors, we used the cryogenic insert equipped with two temperature sensors and the heater. The first thermometer was calibrated silicon diode DT-670B-CU (LakeShore). The second thermometer was the carbon composite resistor Allen Bradly. The carbon resistor was used to measure the temperature in the magnetic field and calibrated by using the silicon diode mounted into the cryogenic insert.

Table 1.

|  | h, nm | $R_{s300}$, Ω/sq | $R_{s300}/R_{s20}$ | $T_c$, K |
|---|---|---|---|---|
| 654 | 9 | 450 | 0.83 | 9.3 |
| 682 | 8 | 525 | 0.73 | 8.95 |
| 803 | 7 | 520 | 0.64 | 8.1 |
| 1181 | 3.5 | 1260 | 0.56 | 5.05 |
| 1180 | 2.5 | 1820 | 0.49 | 3.55 |
| 1179 | 2 | 4410 | 0.13 | --- |

The resistance heater allows fine adjustment of the temperature. In order to measure the temperature dependence of the film surface resistance in the wide range of temperatures, we mounted the cryogenic insert in our standard double-walled vacuum-insulated dipstick [31] with the minimum temperature of 1.6 K. During measurement the temperature dependences of resistance in a magnetic field, the cryogenic insert in simple single-wall dipstick was placed into a superconducting solenoid with the maximum magnetic induction of 1.58 T at the constant current 44.6 A. To ensure that there was no possible difference between the temperature of the investigated sample and the thermometer, the measurements were carried out at decreasing and increasing temperatures. The agreement of the obtained results proves the validity of the temperature measurements.

In order to measure the quantum efficiency, the detectors were installed in a special holder. The installation technique of the detector and its coupling with a single-mode optical fiber were described earlier [12]. The calculated optical coupling between the detector active area and the radiation propagating in the fiber core is more than 99 % at wavelength 1310 nm taking into account the data for the mode field diameter, the Gaussian distribution in single-mode fiber, and possible misalignment of 1 μm. The detectors in the holders were mounted in a custom made cryostat based on a closed cycle refrigerator (SUMITOMO, RDK-101D) which provides an operation temperature of less than 2.3K. The electrical output signal and the detector bias current were transferred through CuNi coaxial cables and hermetically sealed SMA connectors. The input optical signal was send through a single-mode fiber SMF28e and a custom made hermetically sealed input connector which provides insertion losses of less than 0.1 dB.

To determine the quantum efficiency, we used a standard technique for measuring the input in-fiber radiation power at a particular wavelength and the number of voltage pulses appearing on the detector during absorption of this radiation. Thus, the measurements results of the quantum efficiency showed below represent the measurement of the system quantum efficiency (SQE) which includes both the optical coupling losses between the detector and the single-mode fiber and the insertion losses in the hermetically sealed fiber feedthrough. The measurement setup includes the following equipment: the laser diode used as a source of radiation (Dual Laser Source FHS1D02, radiation wavelength $\lambda$ = 1310 nm), the calibrated power meter Ophir PD300-IRG-V1, the calibrated fiber optic attenuator EXPO FVA-600, the fiber polarization controller, the detector bias source integrated with the voltage amplifier system (standard SCONTEL's Control Unit), the pulse counter Agilent 53131A and the oscilloscope Tektronix DPO 71604. The overall relative error in the system quantum efficiency measurements is determined by the accuracy of the power measurements and is 3 %.

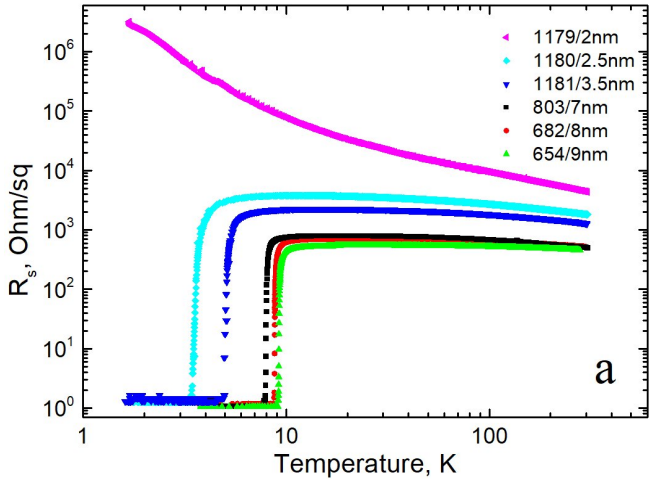 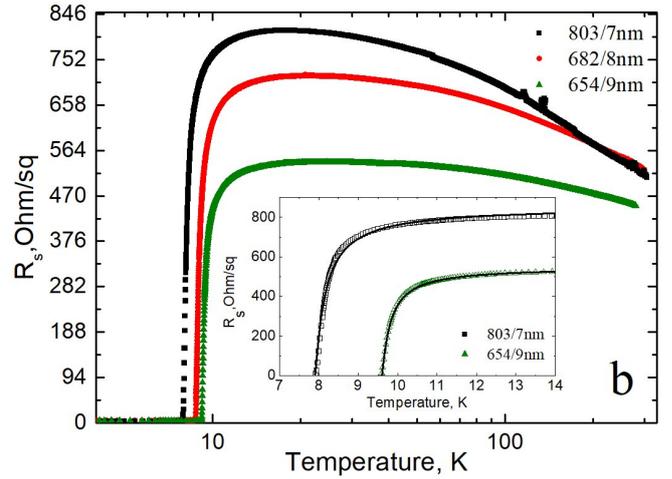

Figure 1. The experimental dependences $R_s(T)$ for NbN films: a) 654, 682, 803, 1181, 1180, 1179; b) 654, 682, 803 (scaling). Insert on b) shows the experimental and the calculated plot of the dependence $R_s(T)$ for the film 654.

The electron diffusion coefficient was determined directly in SSPD detectors using a standard method for the measurement of the dependence of superconducting transition temperature versus magnetic field [32, 33] as follows:

$$D = -\frac{4k_B}{\pi e} \cdot \left[\frac{dB}{dT_C}\right]^{-1}, \quad (1)$$

where $k_B$ is the Boltzmann constant, $e$ is the charge of electron, and $dB/dT_C$ is the derivative of the magnetic field with respect to superconducting transition temperature.

**Study of the $R_s(T)$ dependences for NbN films.** One of the main purposes of the current work is the investigation of NbN thin films with different thicknesses and RRRs for fabrication of SSPD detectors which have saturated quantum efficiency in the range of bias currents close to critical current $I_c$. We tried to fix the other parameters of detectors which could also have influence on quantum efficiency. These fixed parameters are the fabrication technology of structures, the planar topology of superconducting single-photon detectors, the photon wavelength, and the detector operation temperature. The varying parameter are the thickness ($h$) of superconducting NbN film, the film surface resistance $R_s$ at $T = 300$ K ($R_{s300}$) and at a temperature close to the superconducting transition $T = 20$ K ($R_{s20}$), and also the superconducting transition temperature $T_C$. The above characteristics for six solid NbN films are shown in Table 1.

The coefficient $R_{s300}/R_{s20}$ also decreases with decreasing thickness and compound of superconducting films. It should be noted that the value of $R_{s20}$ for all films, except for 1179, approximately corresponds to the maximum value of the resistance for the dependence $R_s(T)$. The resistance for the thinnest NbN film (1179) increases continuously with decreasing temperature below 20 K. In this case, the resistance $R_{s20}$ for the film 1179 is almost one order of magnitude higher than the value of $R_{s300}$. Also, this film does not show the tendency toward decreasing resistance with decreasing temperature down to 1.6 K where the resistance grows by almost three orders of magnitude compared to the value at room temperature.

In order to explain the obtained dependencies of $R_s(T)$ (figure 1), we found that the same conductivity behavior of the quasi-two-dimensional disordered metallic films was observed in other works [34-37]. The authors explained the significant increase of the film resistance with decreasing temperature and the observed superconductor-insulator transition caused by the influence of quantum corrections to the conductivity. Taking account the quantum corrections to the conductivity (the Aslamazov-Larkin correction $\sigma_{AL}$, the correction to the density of states $\sigma_{DOS}$, the Maki-Thompson correction $\sigma_{MT}$; weak localization correction $\sigma_{WL}$, the correction $\sigma_{ID}$) the total surface resistance of the film can be represented as follows [1, 38-44]:

$$R_S = (\sigma_0 + \sigma_{AL} + \sigma_{DOS} + \sigma_{MT} + \sigma_{WL} + \sigma_{ID})^{-1}, \quad (2)$$

where $\sigma_0$ is the surface conductivity of the film in accordance with the Drude theory of metal. The quantum corrections to the conductivity can be found as follows:

$$\sigma_{AL} = \frac{e^2}{16\varepsilon}, \quad \sigma_{DOS} = \frac{e^2}{2\pi^2} \ln \frac{\varepsilon}{\ln \frac{k_B T_C \tau}{\hbar}}, \quad (3)$$

$$\sigma_{MT} = \frac{e^2}{2\pi^2} A\beta \ln \frac{2\pi}{e^2 R_S \ln \frac{\pi}{e^2 R_S}}, \quad (4)$$

$$\sigma_{WL} + \sigma_{ID} = \frac{e^2}{2\pi^2} A_{WL+ID} \ln \frac{k_B T \tau}{\hbar}, \quad (5)$$

Where $\varepsilon = ln(T/T_C)$, $T_C$ is the superconducting transition temperature, $\beta$ is the Larkin function [45], and $\tau$ is the phase relaxation time. The value of $\tau$ and its temperature dependence were taken from the paper [46]:

$$\tau = \frac{2\pi^2}{e^2 R_S k_B T \ln \frac{\pi}{e^2 R_S}}. \quad (6)$$

The coefficient $A_{WL+ID}$ is the fitting parameter which determines mutual influence of $\sigma_{WL}$ and $\sigma_{ID}$ and defines the behaviour of the dependence $R_s(T)$ in the temperature range higher than $T_C$. The critical temperature was taken from the dependence $R_s(T)$ and defined the behaviour $R_s(T)$ in the temperature range near and below the superconducting transition. Also the parameter $A$ was introduced in the Maki-Thompson correction $\sigma_{MT}$ (4) to take into account the fact that according [47, 48] $\beta$ takes values smaller than the tabulated theoretical values shown in [45] near the critical temperature ($T/T_C < 2$). So, we used several fitting parameters but which are independent and can be determined in the only possible way. The inset on the figure 1b shows the comparison of the experimental $R_s(T)$ dependence for the film 654 and the calculated dependence (2) obtained with the following parameters: $T_C = 9.3$ K, $A_{WL+ID} = 0.6$, the film surface resistance 537 Ω/sq under the normal state resistivity at the temperature of $T = 15$ K, coefficient $A = 0.3$. Also we found that the quantum correction $\sigma_{WL}+\sigma_{ID}$ (or the value of $A_{WL+ID}$) increases with decreasing film thickness by adjusting the parameters fitted to experimentally observed dependencies $R_s(T)$ for other investigated films. The absence of superconducting transition for the NbN film 1179 at the temperature down to 1.6 K is probably due to the superconducting transition at the lower temperature or the transition of NbN film into a dielectric state.

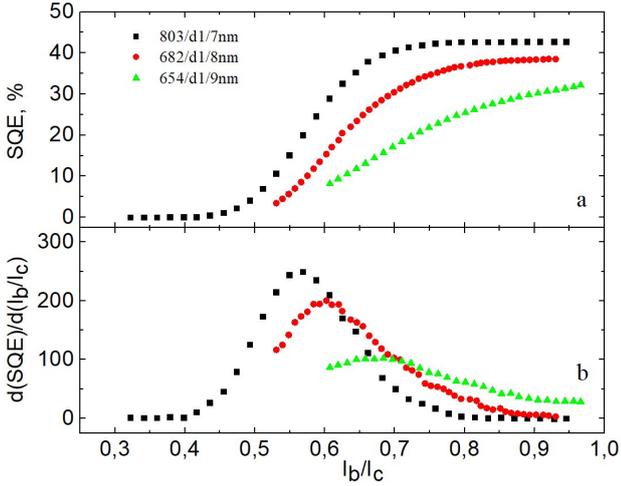

Figure 2. System QE vs. ($I_b/I_c$) dependence measured at the wavelength of 1310 nm a) and the derivative $d(QE)/d(I_b/I_c)$ vs. ($I_b/I_c$) dependence (b) for the detectors 803/d1, 682/d1, 654/d1.

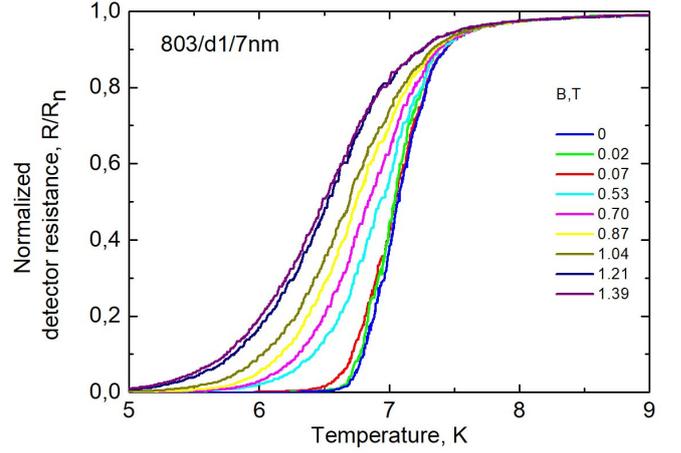

Figure 3. Dependences of resistance vs. temperature in the magnetic field for superconducting single-photon detector 803/d1.

**Measurement of the dependence $QE(I_b/I_c)$.** We fabricated three batches of SSPD detectors based on NbN films 654, 682, 803 and measured its quantum efficiency. The critical current density of detectors at $T = 2.2$ K varies within $1.5 \div 2 \times 10^6$ A/cm$^2$ except for several structures in each batch which have evident constrictions in NbN superconducting strip. Our more detailed research of batch uniformity in terms of resistance, critical current and quantum efficiency can be found elsewhere [49]. Films 1181 and 1180 had significantly low superconducting transition temperature and were not used to fabricate SSPD detectors. Film 1179 did not have a transition into a superconducting state.

It is natural to assume that each photon absorbed by a superconducting structure will give a voltage pulse if the quantum efficiency of a SSPD detector doesn't vary and reaches saturation level with increasing detector bias current. It means that the "internal" quantum efficiency of a detector is close to the unity, while the value of SQE measured during the experiment is determined by the detector absorption coefficient and optical coupling between the superconducting structure and the radiation.

It should be noted that the absorption coefficient for a given wavelength is determined by the thickness of Au/Si$_3$N$_4$ layers which form an optical cavity located under the superconducting structure, the surface resistance of the film, and the meander filling factor. Therefore, the absorption coefficient for structures made of different films may differ slightly even for identical optical cavity. Figure 2a shows the measurement results of the system quantum efficiency vs. the normalized bias current $I_b/I_c$ for three detectors (654/d1, 682/d1, 803/d1) taken from three different batches (654, 682, 803). They had similar values of the system quantum efficiency SQE~40% at bias currents close to $I_c$. These detectors were made on substrate having only Au/Si$_3$N$_4$ sub-layers optical cavity and didn't have an anti-reflection coating on top the detector. Figure 2b shows the dependencies $d(SQE)/d(I_b/I_c)$ vs. $I_b/I_c$ for these detectors which support the understanding of the behaviour of the quantum efficiency as function of the normalized bias current.

The shown curves demonstrate the following features:

1. Despite the close values of the quantum efficiency in the range of currents close to the critical one, the dependencies $SQE(I_b/I_c)$ differ significantly for the three detectors. The dependence $SQE(I_b/I_c)$ for detector 803/d1 reaches a constant value of SQE (dependence with saturation) in the range $I_b/I_c > 0.8$ which is clearly demonstrated in Fig. 2b where the derivative $d(SQE)/d(I_b/I_c)$ turns into zero at the indicated values of the current. The dependence $SQE(I_b/I_c)$ for the detector 682/d1 approaches saturation values only at the largest bias currents and the derivative $d(SQE)/d(I_b/I_c)$ for this detector turns into zero at the currents $I_b/I_c$ close to 0.95. The dependence $SQE(I_b/I_c)$ for the detector 654/d1 demonstrates the monotonic growth with increasing current and $d(SQE)/d(I_b/I_c)$ does not achieve zero.

2. For detectors whose $SQE(I_b/I_c)$ dependence demonstrate more obvious tendency to saturation maximum values of $d(SQE)/d(I_b/I_c)$ are higher and the values of $I_b/I_c$ at which the derivative of the quantum efficiency achieves its maximum value are lower. Also it can be noted that the $SQE(I_b/I_c)$ and $d(SQE)/d(I_b/I_c)$ dependencies for detector 803/d1 are quite symmetrical with respect to the current $I_b/I_c$ corresponding to the maximum value of $d(SQE)/d(I_b/I_c)$.

From the comparison of the results presented in figure 1 and figure 2, it can be seen that the saturation of the quantum efficiency in the range of high currents is more clearly pronounced for detectors made from thinner NbN films characterized by lower values of $R_{s300}/R_{s20}$, a smaller value of the superconducting transition temperature. Since the quantum corrections to conductivity play an important role in the conductivity of thinner films, as shown above, it can be concluded that the tendency to saturation of the dependence $QE(I_b/I_c)$ can be explained by the disorder of superconducting films.

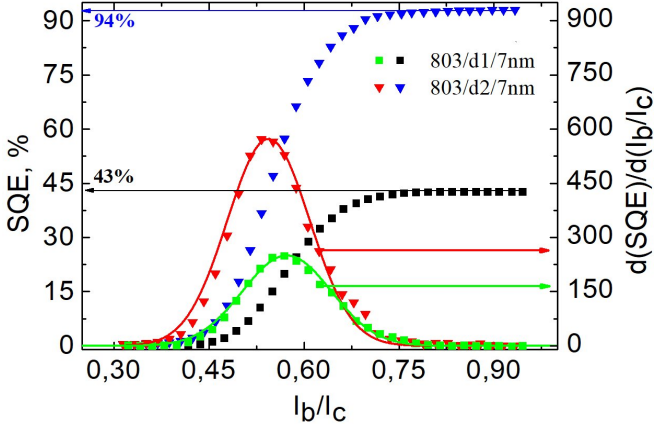

Figure 4. System $QE$ vs $(I_b/I_c)$ dependences measured at the wavelength of 1310 nm a) and the derivative $d(QE)/d(I_b/I_c)$ vs. $(I_b/I_c)$ (b) dependences for detectors 803/d1 and 803/d2 produced on the basis of the NbN film 803. Solid lines are the best fit of Gaussian distribution.

It is natural to assume that the increase in the degree of disorder for NbN films structure and NbN films surface resistance are associated with decreasing the electron diffusion constant $D$. And this assumption was confirmed experimentally. The measured values of the diffusivity were 0.63 cm$^2$/s and 0.74 cm$^2$/s for the samples 803/d1 and 654/d1, respectively. Using these values, we estimate the Ioffe–Regel parameter $k_Fl=3Dm/\hbar$ of 1.6 and 1.9 for 7 nm (sample 803/d1) and 9 nm (sample 654/d1) thick NbN films, respectively, where m is the rest mass of the electron. These parameters $k_Fl$ are in good agreement with the value found in [50] for strongly disordered film (1.3 for 10 nm thick NbN). The similar result were presented in [51] for WSi film with the diffusion coefficient of 0.75 cm$^2$/s which gives $k_Fl$ of 1.9. Additionally, the specific behaviour of $R(T)$ for both samples in a magnetic field (figure 3 presents the graph only for 654/d1) is in good agreement with the typical dependencies for samples where the quantum corrections to the conductivity exhibit considerable influence [52]. Thus, $R(T)$ curves not be parallel shifted in the direction of lower temperature with increasing of magnetic field induction which is typical for sufficiently thick superconducting films. However, $R(T)$ curves demonstrate a significant increase of the width of the superconducting transition by shifting its low-temperature tail.

When the value of $D$ decreases, the probability of the appearance of a voltage pulse on a superconducting structure after the absorption of a photon should increase for two reasons [53]: a) At the initial stage of hot spot formation the possibility of the thermalization of hot electrons by means of their diffusion is reduced. b) The time of the inelastic electron-electron interaction $\tau_{e-e}$ decreases which leads to decreasing thermalization time of electrons due to their interaction and decreasing superconductor area over which the energy of the absorbed photon is distributed. Thus, it leads to greater influence on its superconducting properties and increases the probability of photon detection. The difference in the behaviour of dependencies of the quantum efficiency vs. normalized bias current is associated with a change in the structural ordering of initial superconducting NbN films and a change of their diffusivity.

As mentioned above, three batches of SSPD detectors based on NbN films (654, 682, 803) were fabricated. The detectors with only Au/Si$_3$N$_4$ bilayer cavity were used for the comparison and have the values of the system quantum efficiency of ~40-50 %. At the same time, on top of the five detectors (7 nm thick film) were made additional antireflection coating consist Al$_2$O$_3$/Si/Al$_2$O$_3$ layers which significantly increase the absorption coefficient of the structure. SQE for all five samples was in the range of 88-94%. The best value of SQE for detector 803/d2 was measured as high as 94±3% at the wavelength of 1310 nm for optimal polarization which is very close to recently reported SQE over 90% for 1310 nm and 1550 nm [54, 55]. The indicated accuracy is determined by the calibration error of the input photon number. The $SQE(I_b/I_c)$ and $d(SQE)/d(I_b/I_c)$ dependencies for detectors 803/d2 and 803/d1 fabricated from same film are shown in figure 4. The presented dependencies demonstrate the full correspondence of the main features, except for the achieved constant value of the quantum efficiency at bias currents $I_b > 0.8I_c$ which confirms that the deposition process of multilayer anti-reflection coating does not affect the properties of a NbN film.

In addition, the symmetry of the $SQE(I_b/I_c)$ and $d(SQE)/d(I_b/I_c)$ dependencies with respect to $I_b/I_c$ value corresponding to the maximum of $d(SQE)/d(I_b/I_c)$, the qualitative agreement of the $SQE(I_b/I_c)$ dependence with a standard probability-distribution function, and good approximation of the $d(SQE)/d(I_b/I_c)$ dependence with the Gaussian distribution (figure 4) confirm the fact that the probability of appearance of a voltage pulse in a SSPD after the absorption of photon is determined by many random and independent events such as, for example, the place where the photon was absorbed across the strip width.

The analysis of the Gaussian dependencies which describe the $d(SQE)/d(I_b/I_c)$ curves also shows that the mathematical expectation for these dependencies (the value of the current $I_b/I_c$ corresponding to the maximum value of the derivative $d(SQE)/d(I_b/I_c)$ and the mean-square deviation for these dependencies which is 0.065, 0.07, 0.082, 0.140 for the samples 803/d2, 803/d1, 682/d1, 654/d1, respectively, can also be used to estimate the quality of the fabricated detectors. The ability to reach 100% of the internal quantum efficiency is increased for the detectors with smaller values of both mathematical expectation and mean-square deviation.

**Conclusions**

We have investigated the influence of the thickness of superconducting NbN film, as well as the dependent values of $R_{s300}/R_{s20}$ on the tendency toward to saturation of $SQE(I_b/I_c)$ dependencies for superconducting single-photon detectors in the range of currents close to $I_c$. We have shown that the decrease of $R_{s300}/R_{s20}$ and the decrease of the electron diffusivity with decreasing NbN film thickness are associated with increasing influence of the quantum corrections to conductivity. The fabrication of SSPD detectors based on structurally disordered films allows to reach saturated dependence $SQE(I_b/I_c)$ in the range of currents > $0.8I_b/I_c$. The system quantum efficiency for the best-investigated detector with ARC approaches ultimate value of 94±3% at wavelength of 1310 nm.

**References**

[1] Maki K 1968 The critical fluctuation of the order parameter in type-II superconductors *Progress of Theoretical Physics* **39** 897-906
[2] Marsili F, Verma V B, Stern J A, Harrington S, Lita A E, Gerrits T, Vayshenker I, Baek B, Shaw M D and Mirin R P 2013 Detecting single infrared photons with 93% system efficiency *Nature Photonics* **7** 210-4
[3] Rosenberg D, Kerman A, Molnar R and Dauler E 2013 High-speed and high-efficiency superconducting nanowire single photon detector array *Optics express* **21** 1440-7
[4] Verma V B, Korzh B, Bussieres F, Horansky R D, Dyer S D, Lita A E, Vayshenker I, Marsili F, Shaw M D and Zbinden H 2015 High-efficiency superconducting nanowire single-photon detectors fabricated from MoSi thin-films *Optics express* **23** 33792-801
[5] Zhang W, Li H, You L, Huang J, He Y, Zhang L, Liu X, Chen S, Wang Z and Xie X 2016 Superconducting Nanowire Single-Photon Detector With a System Detection Efficiency Over 80% at 940-nm Wavelength *IEEE Photonics Journal* **8** 1-8
[6] Shibata H, Shimizu K, Takesue H and Tokura Y 2015 Ultimate low system dark-count rate for superconducting nanowire single-photon detector *Optics letters* **40** 3428-31
[7] Smirnov K, Vachtomin Y, Divochiy A, Antipov A and Goltsman G 2015 Dependence of dark count rates in superconducting single photon detectors on the filtering effect of standard single mode optical fibers *Applied Physics Express* **8** 022501
[8] Yang X, Li H, Zhang W, You L, Zhang L, Liu X, Wang Z, Peng W, Xie X and Jiang M 2014 Superconducting nanowire single photon detector with on-chip bandpass filter *Optics express* **22** 16267-72
[9] Shcheslavskiy V, Morozov P, Divochiy A, Vakhtomin Y, Smirnov K and Becker W 2016 Ultrafast time measurements by time-correlated single photon counting coupled with superconducting single photon detector *Review of scientific instruments* **87** 053117
[10] Wu J, You L, Chen S, Li H, He Y, Lv C, Wang Z and Xie X 2017 Improving the timing jitter of a superconducting nanowire single-photon detection system *Applied Optics* **56** 2195-200
[11] Kerman A J, Rosenberg D, Molnar R J and Dauler E A 2013 Readout of superconducting nanowire single-photon detectors at high count rates *Journal of Applied Physics* **113** 144511
[12] Sidorova M V, Divochiy A V, Vakhtomin Y B and Smirnov K V 2015 Ultrafast superconducting single-photon detector with a reduced active area coupled to a tapered lensed single-mode fiber *Journal of Nanophotonics* **9** 093051-
[13] Zhao Q, Jia T, Gu M, Wan C, Zhang L, Xu W, Kang L, Chen J and Wu P 2014 Counting rate enhancements in superconducting nanowire single-photon detectors with improved readout circuits *Optics letters* **39** 1869-72
[14] Seleznev V, Divochiy A, Vakhtomin Y B, Morozov P, Zolotov P, Vasil'ev D, Moiseev K, Malevannaya E and Smirnov K 2016 Superconducting detector of IR single-photons based on thin WSi films. In: *Journal of Physics: Conference Series*: IOP Publishing) p 012032
[15] Gemmell N R, McCarthy A, Liu B, Tanner M G, Dorenbos S D, Zwiller V, Patterson M S, Buller G S, Wilson B C and Hadfield R H 2013 Singlet oxygen luminescence detection with a fiber-coupled superconducting nanowire single-photon detector *Optics express* **21** 5005-13
[16] Grein M, Dauler E, Kerman A, Willis M, Romkey B, Robinson B, Murphy D and Boroson D 2015 A superconducting photon-counting receiver for optical communication from the Moon *SPIE Newsroom* **9**
[17] Vorobyov V V, Kazakov A Y, Soshenko V V, Korneev A A, Shalaginov M Y, Bolshedvorskii S V, Sorokin V N, Divochiy A V, Vakhtomin Y B and Smirnov K V 2017 Superconducting detector for visible and near-infrared quantum emitters *Optical Materials Express* **7** 513-26
[18] Yamamoto J, Oura M, Yamashita T, Miki S, Jin T, Haraguchi T, Hiraoka Y, Terai H and Kinjo M 2015 Rotational diffusion measurements using polarization-dependent fluorescence correlation spectroscopy based on superconducting nanowire single-photon detector *Optics express* **23** 32633-42
[19] Ivry Y, Kim C-S, Dane A E, De Fazio D, McCaughan A N, Sunter K A, Zhao Q and Berggren K K 2014 Universal scaling of the critical temperature for thin films near the superconducting-to-insulating transition *Physical Review B* **90** 214515
[20] Marsili F, Najafi F, Dauler E, Bellei F, Hu X, Csete M, Molnar R J and Berggren K K 2011 Single-photon detectors based on ultranarrow superconducting nanowires *Nano letters* **11** 2048-53
[21] Najafi F, Marsili F, Dauler E, Molnar R and Berggren K 2012 Timing performance of 30-nm-wide superconducting nanowire avalanche photodetectors *Applied Physics Letters* **100** 152602
[22] Korneeva Y P, Mikhailov M Y, Pershin Y P, Manova N, Divochiy A, Vakhtomin Y B, Korneev A, Smirnov K, Sivakov A and Devizenko A Y 2014 Superconducting single-photon detector made of MoSi film *Superconductor Science and Technology* **27** 095012
[23] Verma V B, Korzh B, Bussières F, Horansky R D, Lita A E, Marsili F, Shaw M, Zbinden H, Mirin R and Nam S 2014 High-efficiency WSi superconducting nanowire single-photon detectors operating at 2.5 K *Applied Physics Letters* **105** 122601
[24] Tanner M G, Natarajan C, Pottapenjara V, O'Connor J, Warburton R, Hadfield R, Baek B, Nam S, Dorenbos S and Ureña E B 2010 Enhanced telecom wavelength single-photon detection with NbTiN superconducting nanowires on oxidized silicon *Applied Physics Letters* **96** 221109
[25] Pernice W H, Schuck C, Minaeva O, Li M, Goltsman G, Sergienko A and Tang H 2012 High-speed and high-efficiency travelling wave single-photon detectors embedded in nanophotonic circuits *Nature communications* **3** 1325
[26] Lipatov A, Okunev O, Smirnov K, Chulkova G, Korneev A, Kouminov P, Gol'tsman G, Zhang J, Slysz W and Verevkin A 2002 An ultrafast NbN hot-electron single-photon detector for electronic applications *Superconductor Science and Technology* **15** 1689
[27] Lusche R, Semenov A, Ilin K, Siegel M, Korneeva Y, Trifonov A, Korneev A, Goltsman G, Vodolazov D and Hübers H-W 2014 Effect of the wire width on the intrinsic detection efficiency of superconducting-nanowire single-photon detectors *Journal of Applied Physics* **116** 043906
[28] Smirnov K, Divochiy A, Vakhtomin Y B, Sidorova M, Karpova U, Morozov P, Seleznev V, Zotova A and Vodolazov D Y 2016 Rise time of voltage pulses in NbN superconducting single photon detectors *Applied Physics Letters* **109** 052601
[29] Gol'Tsman G, Smirnov K, Kouminov P, Voronov B, Kaurova N, Drakinsky V, Zhang J, Verevkin A and Sobolewski R 2003 Fabrication of nanostructured superconducting single-

photon detectors *IEEE Transactions On Applied Superconductivity* **13** 192-5
[30] Rosfjord K M, Yang J K, Dauler E A, Kerman A J, Anant V, Voronov B M, Gol'Tsman G N and Berggren K K 2006 Nanowire single-photon detector with an integrated optical cavity and anti-reflection coating *Optics Express* **14** 527-34
[31] Smirnov K, Vachtomina Y B, Ozhegova R, Pentina I, Slivinskayab E, Korneeva A and Goltsmana G 2008 Fiber coupled single photon receivers based on superconducting detectors for quantum communications and quantum cryptography. In: *Proc. of SPIE Vol,* pp 713827-1
[32] Gershenzon E, Gershenzon M, Gol'tsman G, Lyul'kin A, Semenov A and Sergeev A 1990 Electron-phonon interaction in ultrathin Nb films *Sov. Phys. JETP* **70** 505-11
[33] Karasik B, Il'in K, Pechen E and Krasnosvobodtsev S 1996 Diffusion cooling mechanism in a hot-electron NbC microbolometer mixer *Applied physics letters* **68** 2285-7
[34] Haviland D, Liu Y and Goldman A 1989 Onset of superconductivity in the two-dimensional limit *Physical Review Letters* **62** 2180
[35] Strongin M, Thompson R, Kammerer O and Crow J 1970 Destruction of superconductivity in disordered near-monolayer films *Physical Review B* **1** 1078
[36] Vinokur V M, Baturina T I, Fistul M V, Mironov A Y, Baklanov M R and Strunk C 2008 Superinsulator and quantum synchronization *Nature* **452** 613-5
[37] Kang L, Jin B, Liu X, Jia X, Chen J, Ji Z, Xu W, Wu P, Mi S and Pimenov A 2011 Suppression of superconductivity in epitaxial NbN ultrathin films *Journal of Applied Physics* **109** 033908
[38] Al'Tshuler B, Aronov A, Larkin A and Khmel'Nitskii D 1981 Anomalous magnetoresistance in semiconductors *Sov. Phys. JETP* **54** 411-9
[39] Altshuler B, Varlamov A and Reizer M Y 1983 Interelectron effects and the conductivity of disordered two-dimensional electron systems *ZHURNAL EKSPERIMENTALNOI I TEORETICHESKOI FIZIKI* **84** 2280-9
[40] Altshuler B L, Aronov A G and Lee P 1980 Interaction effects in disordered Fermi systems in two dimensions *Physical Review Letters* **44** 1288
[41] Aslamasov L and Larkin A 1968 The influence of fluctuation pairing of electrons on the conductivity of normal metal *Physics Letters A* **26** 238-9
[42] Aslamazov L and Larkin A 1968 Effect of fluctuations on the properties of a superconductor at temperatures above the critical temperature(Electron fluctuation coupling effect on superconductor kinetic properties at temperature above critical temperature) *Fizika tverdogo tela* **10** 1104-11
[43] Hikami S, Larkin A I and Nagaoka Y 1980 Spin-orbit interaction and magnetoresistance in the two dimensional random system *Progress of Theoretical Physics* **63** 707-10
[44] Thompson R S 1970 Microwave, flux flow, and fluctuation resistance of dirty type-II superconductors *Physical Review B* **1** 327
[45] Larkin A 1980 Reluctance of two-dimensional systems *JETP Lett* **31** 219-23
[46] Efros A L and Pollak M 2012 *Electron-electron interactions in disordered systems* vol 10: Elsevier)
[47] Giannouri M, Papastaikoudis C and Rosenbaum R 1999 Low-temperature transport properties of Nb 1− x Ta x thin films *Physical Review B* **59** 4463
[48] Giannouri M, Rocofyllou E, Papastaikoudis C and Schilling W 1997 Weak-localization, Aslamazov-Larkin, and Maki-Thompson superconducting fluctuation effects in disordered Zr 1− x Rh x films above T c *Physical Review B* **56** 6148
[49] Zolotov P, Divochiy A, Vakhtomin Y B, Morozov P, Seleznev V and Smirnov K 2017 Development of high-effective superconducting single-photon detectors aimed for mid-IR spectrum range. In: *Journal of Physics: Conference Series*: IOP Publishing) p 062037
[50] Hazra D, Tsavdaris N, Mukhtarova A, Jacquemin M, Blanchet F, Albert R, Jebari S, Grimm A, Blanquet E and Mercier F 2017 The role of Coulomb interaction in superconducting NbTiN thin films *arXiv preprint arXiv:1711.04585*
[51] Kozorezov A, Lambert C, Marsili F, Stevens M J, Verma V B, Stern J A, Horansky R, Dyer S, Duff S and Pappas D P 2015 Quasiparticle recombination in hotspots in superconducting current-carrying nanowires *Physical Review B* **92** 064504
[52] Gantmakher V F and Dolgopolov V T 2010 Superconductor–insulator quantum phase transition *Physics-Uspekhi* **53** 1-49
[53] Vodolazov D Y 2017 Single-Photon Detection by a Dirty Current-Carrying Superconducting Strip Based on the Kinetic-Equation Approach *Physical Review Applied* **7** 034014
[54] Zhang W, You L, Li H, Huang J, Lv C, Zhang L, Liu X, Wu J, Wang Z and Xie X 2017 NbN superconducting nanowire single photon detector with efficiency over 90% at 1550 nm wavelength operational at compact cryocooler temperature *Science China Physics, Mechanics & Astronomy* **60** 120314
[55] Esmaeil Zadeh I, Los J W, Gourgues R B, Steinmetz V, Bulgarini G, Dobrovolskiy S M, Zwiller V and Dorenbos S N 2017 Single-photon detectors combining high efficiency, high detection rates, and ultra-high timing resolution *APL Photonics* **2** 111301